\documentclass[aps,prc,groupedaddress,showpacs,manuscript]{revtex4}

\usepackage{graphicx}

\begin{document}

\title{Transport model analysis of particle correlations in relativistic
    heavy ion collisions at femtometer scales}

\author {Qingfeng Li$\, ^{1}$\footnote{Fellow of the Alexander von Humboldt Foundation.}
\email[]{Qi.Li@fias.uni-frankfurt.de}, Marcus Bleicher$\, ^{2}$,
and Horst St\"{o}cker$\, ^{1,2}$}
\address{
1) Frankfurt Institute for Advanced Studies (FIAS), Johann Wolfgang Goethe-Universit\"{a}t, Max-von-Laue-Str.\ 1, D-60438 Frankfurt am Main, Germany\\
2) Institut f\"{u}r Theoretische Physik, Johann Wolfgang Goethe-Universit\"{a}t, Max-von-Laue-Str.\ 1, D-60438 Frankfurt am Main, Germany\\
 }

%\date{\today}

\begin{abstract}
The pion source as seen through HBT correlations at RHIC energies
is investigated within the UrQMD approach. We find that the
calculated transverse momentum, centrality, and system size
dependence of the Pratt-HBT radii $R_L$ and $R_S$ are
reasonably well in line with experimental data. The predicted
$R_O$ values in central heavy ion collisions are larger as
compared to experimental data. The corresponding quantity
$\sqrt{R_O^{2}-R_S^{2}}$ of the pion emission source is somewhat
larger than experimental estimates.
\end{abstract}

%\textbf{Keywords}:

 \maketitle

In the quest to discover the high temperature phase of Quantum
Chromodynamics (QCD), the Quark Gluon Plasma (QGP),  the beam
energies of accelerators have been boosted upwards from SIS, AGS,
SPS, to RHIC. However, it is well known that the phase transition
from hadrons to quarks might only occur in a small volume part of
the system and within a  rather short timespan in heavy ion
collisions (HICs). This implies that the QGP drops formed at RHIC
might be represented only by a few, locally thermally equilibrated
drops of matter, in which quarks and gluons are de-confined. Thus,
it is essential to probe the space-time structure of the
(equilibrated?) source -- the "region of homogeneity".
Unfortunately, the small size and transient nature of the reactions
preclude direct measurement of the time and/or position. Instead,
correlations of two final-state particles at small relative momenta
provide the most direct link to the size and lifetime of subatomic
reactions. A well-established technique, the  so-called "femtoscopy"
or "HBT" in the heavy-ion community (in reference to Hanbury-Brown
and Twiss's original work with photons) has been extensively used
for HICs with energies from SIS, AGS, SPS, to RHIC
\cite{Lisa:2003ze,Lisa:2000ip,Humanic:2005ye,Frodermann:2006sp,Adams:2003ra,Bearden:1998aq,Adamova:2002qx,Appelshauser:2004ys,Adamova:2002ff,Adamova:2002wi,Chung:2002vk,Ahle:2002mi,Kolb:2003dz,Tomasik:2002rx,Chajecki:2005iv,Nouicer:2005bg,Lisa:2005js,Adler:2004rq,Back:2004ug,Adams:2004yc,Adler:2001zd,Adcox:2002uc,Lisa:2005dd,Heinz:2002un,Pratt:2005bt,Lisa:2000hw,Kniege:2004pt,Csorgo:2004id,Soff:2000eh,Soff:2001hc,Zschiesche:2001dx}.

Numerous HBT-measurements with various two-particle species have
been pursued (see e.g. \cite{Lisa:2005dd,Lisa:2005js} and references
therein). Identically charged pion interferometry has been most
extensively investigated. Basic, but important systematics of
femtoscopic measurements from the AGS, SPS, and RHIC have been
discovered
\cite{Lisa:2005dd,Lisa:2005js,Chajecki:2005iv,Adamova:2002ff}, such
as the dependence  of the HBT radii on system size, collision
centrality, rapidity, transverse momentum, and  particle mass.
However, the existence of the so-called HBT-puzzle (i.e., the fact
that model calculations that incorporate a phase transition to a new
state of matter with many degrees of freedom significantly
over-predict the observed source sizes)
\cite{Lisa:2005dd,Frodermann:2006sp,Soff:2000eh,Soff:2001hc,Zschiesche:2001dx,Heinz:2002un,Pratt:2005bt,Rischke:1995cm,Rischke:1996em}
drives us to a deeper and more systematical theoretical exploration.
The Ultra-relativistic Quantum Molecular Dynamics (UrQMD, v2.2)
transport model (employing hadronic and string degrees of freedom)
(for details, the reader is referred to Refs.\cite{Bass98,
Bleicher99,Bra04,Zhu:2005qa}) and the analyzing program CRAB
(v3.0$\beta$) \cite{Koonin:1977fh, Pratt:1994uf,Pratthome} are
employed here as tools to analyze the two-particle interferometry.
It is known that the probable 
QGP phase is not strictly treated in the present version of UrQMD model for the early-stage 
HICs. However, the failure of fluid dynamical models to get the
$k_T$-dependence of the HBT radii \cite{Lisa:2005dd} suggests that flow is not the
only aspect that influences the observed $k_T$-dependence of HBT radii. A
realistic hadronic cascade model such as UrQMD can thus throw light on what
other mechanisms during the late freeze-out stage induce strong coordinate-momentum 
correlations and thereby generate the observed strong $k_T$-dependence of
the HBT radii. With this equipment, the excitation functions of the HBT radii of
negatively charged pions are calculated systematically. In this
paper, we focus on RHIC energies, where the biggest challenge is
faced by the current theoretical models. The femtoscopy results at
lower energies will be presented in a further study \cite{lqfsoon}.

The correlation function of two particles is decomposed in
Pratt's (so-called longitudinal co-moving system
"Out-Side-Long") three-dimensional convention (Pratt-radii). The three-dimensional
correlation function is fit with the standard Gaussian form:

\begin{equation}
C(q_O,q_S,q_L)=1+\lambda
{\rm exp}(-R_L^2q_L^2-R_O^2q_O^2-R_S^2q_S^2-2R_{OL}^2q_Oq_L), \label{fit1}
\end{equation}
in which $q_i$ and $R_i$ are the components of the pair momentum
difference $\bf{q}=\bf{p}_2-\bf{p}_1$ and the homogeneity length
(Pratt-radii) in the $i$ direction, respectively. The $\lambda$ is
the incoherence factor, which lies between 0 (complete coherence)
and 1 (complete incoherence) in realistic HICs. $R_{OL}$ represents
the cross-term. For mass-symmetric colliding system, 
the $R_{OL}$ vanishes automatically at mid-rapidity due to the longitudinal reflection symmetry  and  is
also found to be negligible in our present calculations.
Furthermore, in the present UrQMD calculations at RHIC energies, the
Coulomb and other potential interactions are not considered (the
"cascade mode" is used) due to the excessive computing times which
would have been used otherwise. The Coulomb final state interactions
are not taken into account in the analyzing program CRAB.

\begin{figure}
\includegraphics[angle=0,width=0.8\textwidth]{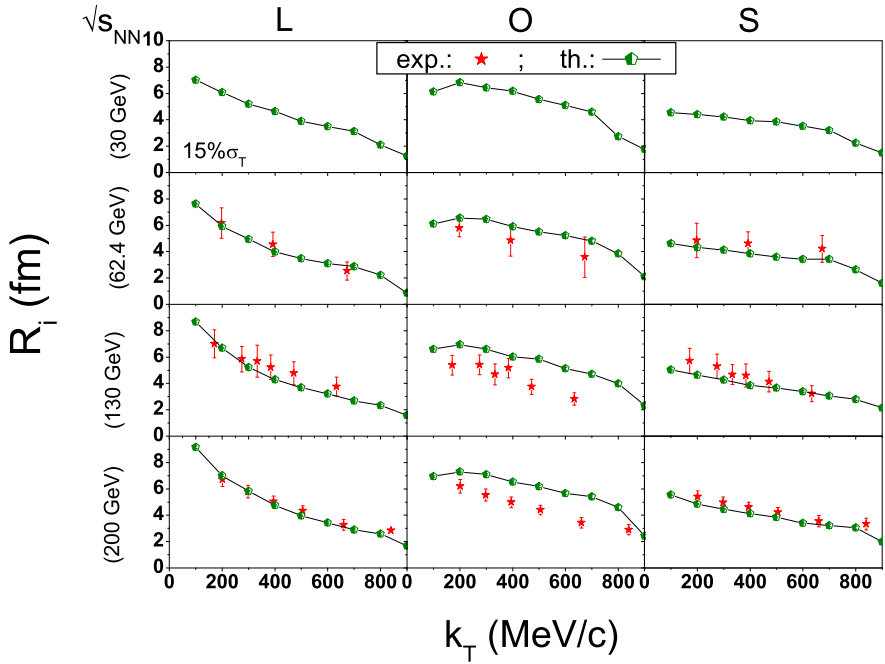}
\caption{(Color online) Transverse momentum $k_T$ dependence (at midrapidity) of the Pratt-radii $R_L$, $R_O$, and $R_S$in Au+Au
collisions at $\sqrt{s_{NN}}=30$, $62.4$, $130$, and $200$ GeV.
Experimental data for the latter three cases are also shown
\cite{Back:2004ug,Adler:2001zd,Adcox:2002uc,Adler:2004rq,Adams:2004yc}.
The experimental errors are the sums of both statistical and systematic
errors. } \label{fig1}
\end{figure}

Fig.\ \ref{fig1} gives the transverse momentum $k_T$ dependence
($\textbf{k}_T=(\textbf{p}_{1T}+\textbf{p}_{2T})/2$) of the
Pratt-radii $R_L$ (left plots), $R_O$ (middle plots), and $R_S$
(right plots) at nucleon-nucleon center-of-mass energies
$\sqrt{s_{NN}}=30$, $62.4$, $130$, and $200$ GeV (plots from top to
bottom) in Au+Au reactions. The experimental results at energies
$\sqrt{s_{NN}}=62.4$, $130$, and $200$ GeV for central collisions
($<15\%$, $<10\%$, and $<5\%$ of the total cross section $\sigma_T$,
respectively) and at mid-rapidities ($|\eta_{cm}|<0.5$) are taken
from Refs.\
\cite{Back:2004ug,Adler:2001zd,Adcox:2002uc,Adler:2004rq,Adams:2004yc}.
The experimental error bars are shown as the sum of both statistical
and systematic errors. The corresponding calculations with the same
trigger- and acceptance- conditions as in the experiments are shown,
as well as the lower energy case $\sqrt{s_{NN}}=30$ GeV for central
collisions ($<15\%$ of $\sigma_T$).

Both the absolute values and the decrease of the Pratt-radii $R_L$
and $R_S$   with transverse momentum is reproduced by the present
model calculations very well. The origin of the decrease of the
Pratt-radii with the increase of transverse momentum is still under
discussion: it may be caused by the strong underlying transverse
flow \cite{Adams:2004yc}, or, by the temperature inhomogeneities
within the hadron source (point of view of the hydrodynamics model)
\cite{Csorgo:2004id}. Here, it is also seen that the calculated
$k_T$-dependence of  $R_S$ is somewhat flatter than that of $R_L$,
which implies that flow effects on the $k_T$-dependence of the
Pratt-radii can at least not be excluded. Besides the flow effect,
the surface-like emission charactistic of microscopic models should
play significantly (or even dominant) role on HBT parameters as well
because also other Cascade/Boltzmann calculations (see e.g., the
Relativistic Quantum Molecular Dynamics model (RQMD)
\cite{Lisa:2000hw,Lisa:2005dd}, the Hadronic Rescattering Model
(HRM) \cite{Humanic:2005ye}, and A Multi-Phase Transport model
(AMPT) \cite{Lin:2002gc}) can reproduce the $k_T$ dependence of
Pratt radii (almost) equally well. The transport of the hadronic
gas, or the final state hadronic 'corona' dynamics, should be
further investigated since the QGP image is obscured by the hadronic
'corona'
\cite{Soff:2000eh,Hirano:2005wx,Werner:2006wp,Werner:2006dx}.   The
UrQMD calculations of $R_L$ and $R_S$ reproduce the experimental
data well within the error bars, while the calculated $R_O$'s are
larger than the experimental data
--- the $R_O$  is about $25\%$ too large. We must conclude that at
RHIC, larger ratios of $R_O$ and $R_S$ are seen from hadron
transport model than expected. Similar observations have also been
reported from most of other model calculations (c.f.
\cite{Lisa:2005dd,Soff:2000eh,Soff:2001hc,Molnar:2002bz,Lin:2002gc,Hirano:2002ds,Zschiesche:2001dx,Soff:2000eh,Humanic:2005ye}).
Ref. \cite{Pratt:2005bt} has argued that the origin of this
HBT-puzzle might be multifaceted.

\begin{figure}
\includegraphics[angle=0,width=0.8\textwidth]{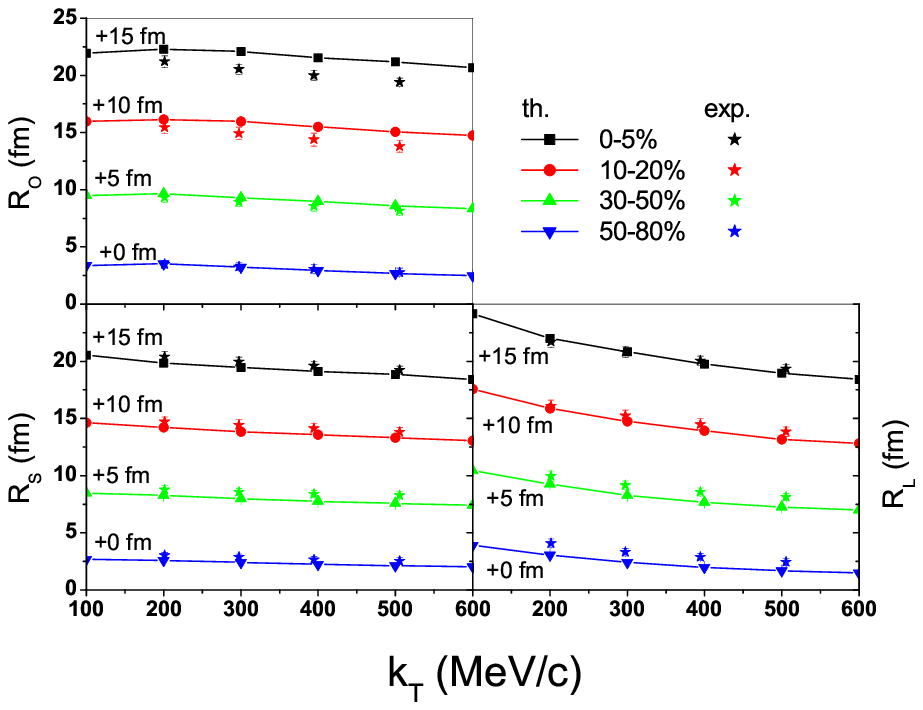}
\caption{(Color online) Midrapidity $k_t$-dependence of Pratt-radii in Au+Au reaction at
$\sqrt{s_{NN}}=200$ GeV for four centralities: $0-5\%$, $10-20\%$,
$30-50\%$, and $50-80\%$ of the total cross section, which are shifted
by 15, 10, 5, and 0 fm, respectively. Experimental data are taken
from Ref. \cite{Adams:2004yc}.} \label{fig2}
\end{figure}

Fig.\ \ref{fig2} shows the $k_T$-dependence of the Pratt-radii in
Au+Au reaction at $\sqrt{s_{NN}}=200$ GeV for four centralities:
$0-5\%$, $10-20\%$, $30-50\%$, and $50-80\%$ of total cross section.
Here, a pseudo-rapidity cut $|\eta_{cm}|<0.5$ has been chosen. For
better visibility, we have shifted in the figure the values of the
radii by 0, 5, 10, and 15 fm for the four centralities. It is very
interesting to see that our calculations for the centrality
dependence of Pratt-radii are in reasonably good agreement with the
experimental data. The important observation, however, is that the
calculated $R_O$ values tend to deviate from the data for central
reactions by $\approx20\%$, while they agree at midcentral and
peripheral collisions.

\begin{figure}
\includegraphics[angle=0,width=0.6\textwidth]{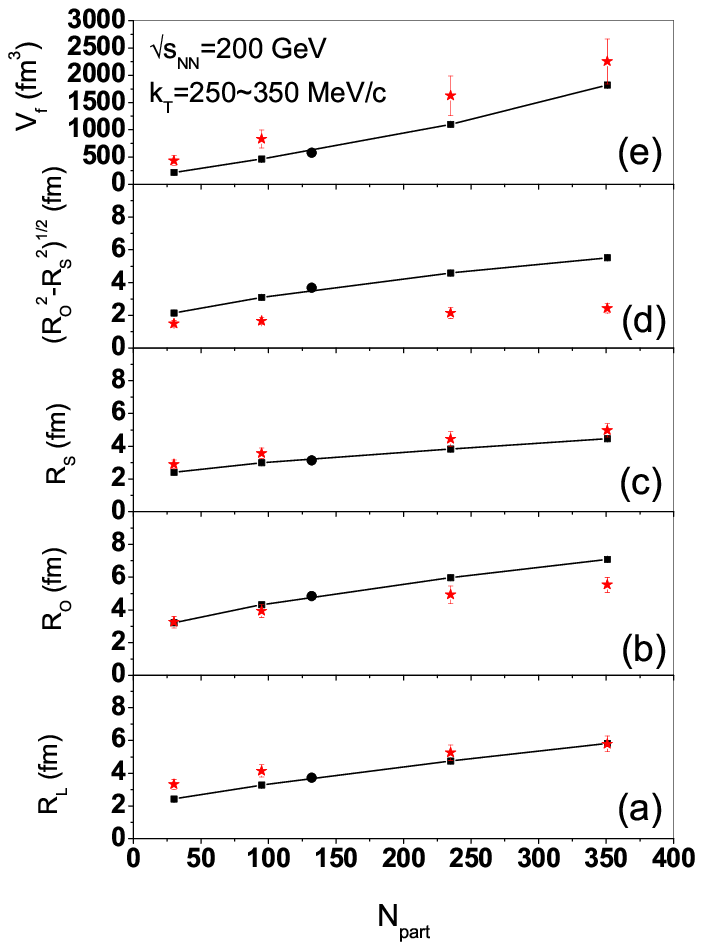}
\caption{(Color online) Centrality dependence of the Pratt-radii ((a)-(c)), the
quantity $\sqrt{R_O^{2}-R_S^{2}}$ (in (d)), and the freeze-out
volume $V_f$ ((e)) at $k_T=250\sim350 {\rm MeV/c}$, in Au+Au collisions
at $\sqrt{s_{NN}}=200$ GeV. Experimental data are taken from Ref.
\cite{Adams:2004yc}. The calculated results for central Cu+Cu
collisions shown with solid dots are perfectly located on the
Au+Au systematic curves.} \label{fig3}
\end{figure}

The centrality dependence of the Pratt-radii can be seen more
clearly from Fig.\ \ref{fig3} , which shows the Pratt-radii at
$k_T=250\sim350 {\rm MeV/c}$ as a function of the number of
participants $N_{part}$. The quantity $\sqrt{R_O^{2}-R_S^{2}}$ is
also shown for comparison. In spite of the reasonable results on the
centrality dependence of the Pratt-radii, the calculated quantity
$\sqrt{R_O^{2}-R_S^{2}}$ obviously deviates from that extracted from
data for the most central collisions: it is about twice as large as
measured by experiments. In previous UrQMD calculations on elliptic
flow at RHIC it was found that only $\sim60\%$ of the observed
elliptic flow is produced \cite{Zhu:2005qa}, which is probably due
to a smaller anisotropy in the pressure gradients in the early stage
of RHIC collisions in the UrQMD simulations compared to
hydrodynamics. Based on Ref.\ \cite{Tomasik:2002rx}, the $R_O$
contains the contributions from temporal extent of the source and
becomes larger with a smaller transverse freeze-out momentum of
particle pairs with a certain duration time. While the expansion has
no effect on the $R_S$. The calculated larger $R_O$ and,
correspondingly, the larger calculated quantity
$\sqrt{R_O^{2}-R_S^{2}}$ in comparison to the data might therefore
be related to the elliptic flow problem in the model.  The pion
freeze-out volume $V_f$ has been investigated thoroughly
experimentally \cite{Adamova:2002ff}. $V_f$ can be expressed as
$V_f=(2\pi)^{3/2}R_LR_S^{2}$. The linear increase of $V_f$ with
$N_{part}$ is expected if the pions freeze out at a constant density
$\rho_f$ at a certain beam energy as observed in Ref.
\cite{Adamova:2002ff} and implied in Ref. \cite{Lisa:2005dd}. It can
be explained reasonably well by the present model, although a
smaller "thermal ellipse" is predicted due to a little shorter $R_L$
and $R_S$ values, as shown.

\begin{figure}
\includegraphics[angle=0,width=0.8\textwidth]{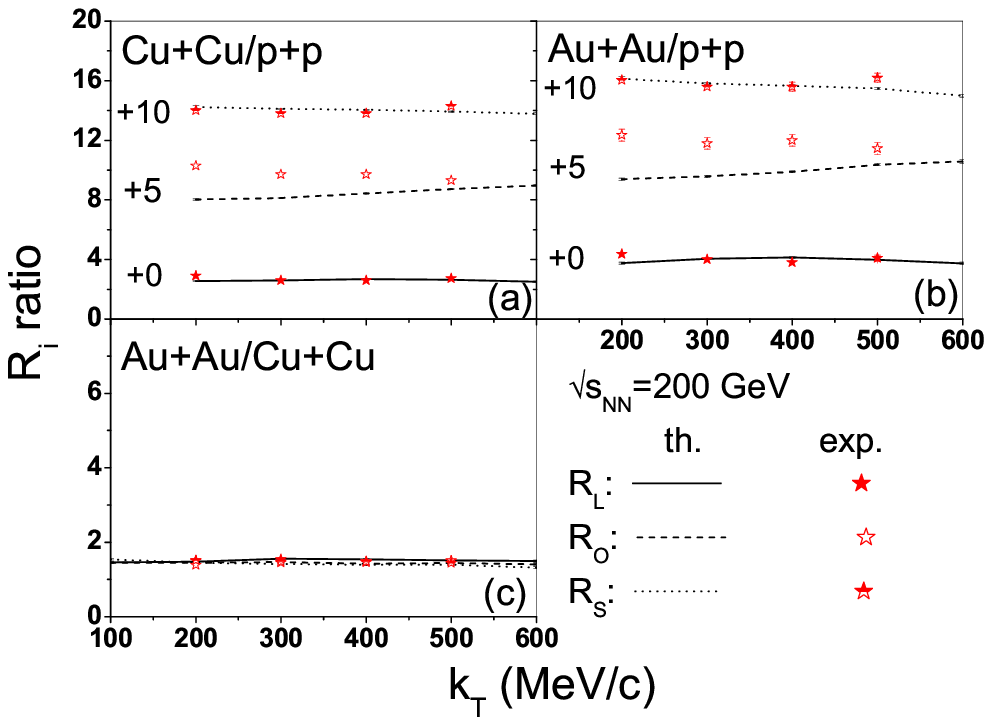}
\caption{(Color online) $k_T$-dependence of the ratios of Pratt-radii between
different systems: (a), Cu+Cu vs. p+p, (b), Au+Au vs. p+p, and (c), Au+Au
vs. Cu+Cu. Central collisions at $\sqrt{s_{NN}}=200$ GeV and a
mid-rapidity cut are chosen. In (a) and (b), the ratios are
shifted by 5 and 10, respectively. The preliminary experimental
data are taken from Ref. \cite{Chajecki:2005iv}.} \label{fig4}
\end{figure}

The calculations for central Cu+Cu collisions are shown in Fig.\
\ref{fig3}  (solid dots), which are in line with the centrality
dependence of the HBT space-time structure calculated for Au+Au
collisions. This implies that the participant  multiplicity is a
very good scaling variable, which drives the geometry (HBT radii) at
mid-rapidity, at least for mid-size to heavy systems. In order to
check this, the $k_T$-dependence of the ratios of the Pratt-radii
between different systems is shown in Fig.\ \ref{fig4} . The radius
ratios shown are from (a), Cu+Cu vs. p+p, (b), Au+Au vs. p+p, and
(c), Au+Au vs. Cu+Cu. In order to read the figure more conveniently,
the $R_O$ and $R_S$ ratios are shifted by 5 and 10, respectively. In
the p+p calculation, the non-femtoscopic correlations at large
relative momenta are also seen, that is, the pion correlations
function saturates at large relative momentum, but the value is not
equal to 1, which was also implied by the preliminary data reported
recently by Ref.\ \cite{Chajecki:2005iv}. We eliminate this effect
by multiplying a constant into the parametrization of the
correlation function. The $R_L$ and $R_S$ values in p+p collisions
can be reproduced well, while the calculated $R_O$ values are again
larger than the experimental data, similar to the nucleus-nucleus
collisions. This might be the origin of the whole puzzle, namely
that the HBT-correlations are somewhat incorrectly put into the
model in the elementary p+p dynamics. Figs.\ \ref{fig4} (a) and (b)
show that the calculated $R_L$ and $R_S$ ratios, reproduce the
experimental data reasonably well. They are almost flat as a
function of $k_T$, which means the $k_T$-dependence of the
Pratt-radii still exists in the elementary p+p collisions at RHIC
energies.

\begin{figure}
\includegraphics[angle=0,width=0.8\textwidth]{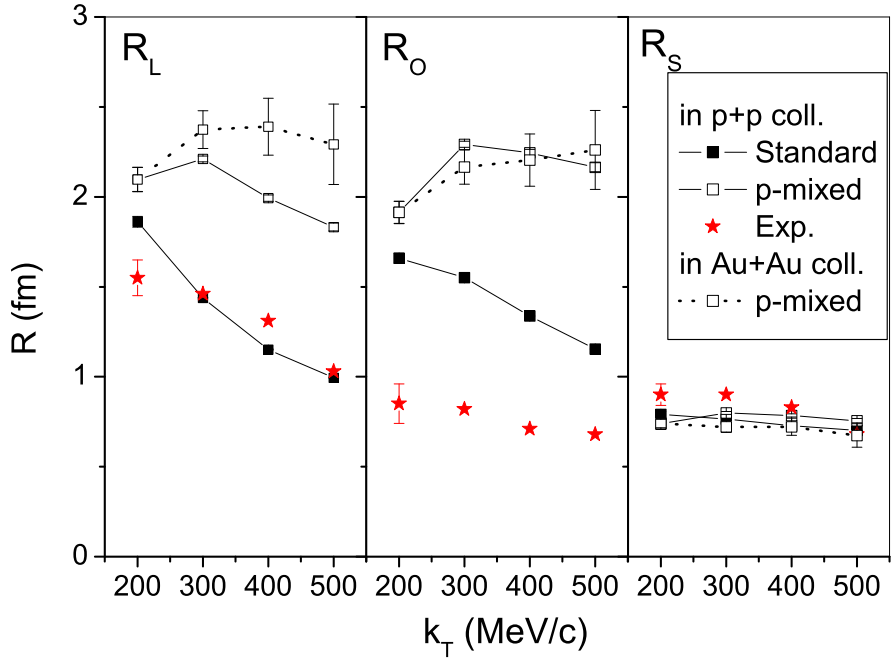}
\caption{(Color online) $k_T$-dependence of the Pratt-radii $R_L$
(left plot), $R_O$ (middle), and $R_S$ (right) in the p+p and the most central Au+Au
collisions at
$\sqrt{s_{NN}}=200$ GeV. The Pratt radii in Au+Au collisions are rescaled by the
Pratt radii at $k_T=200$ MeV$/c$ in the p+p collisions. The standard calculations (titled as
"standard") as well as the calculations after considering the
random mixture of the momenta of freeze-out pions (titled as "p-mixed")
are shown. The preliminary experimental data for p+p collisions are shown with stars. }
\label{fig5}
\end{figure}

The fact that the $k_T$ dependence of $R_L$ and $R_S$ radii in p+p
collisions is similar to AA reaction is puzzling at first glance. In
order to explore the origin of the $k_T$ dependence in pp and AA, we
randomly exchange the momentum vectors of the pions at freeze-out
and recalculated the $k_T$-dependence of Pratt radii in the p+p
(solid lines with open square symbols) and the most central Au+Au
(dotted lines with open square symbols, rescaled by the Pratt radii
at $k_T=200$ MeV$/c$ in the p+p collisions) collisions at RHIC
energy $\sqrt{s_{NN}}=200$ GeV, as shown in Fig.\ \ref{fig5}. We
also show the standard calculation results (solid squares) and the
experimental data (stars) in p+p collisions. After considering the
random mixture of the momenta of freeze-out pions, the $k_T$
dependence of Pratt radii essentially vanishes, especially in the
transverse direction. This is a clear indication that space-momentum
correlations drive the behavior of the Pratt radii with increasing
$k_T$ both in AA and pp! However, it is important to stress that the
origin of the space-momentum correlation in p+p is most probably due
to jet-like structures and not flow.

Since the
UrQMD model gives too large a $R_O$ value in p+p collisions, the
calculated $R_O$-ratio between Au+Au (or Cu+Cu) and p+p is smaller
than the experimental data, in particular at low transverse momenta.
This phenomenon disappears when we consider the Pratt-radii-ratios
between two heavy systems, for examples, between Au+Au and Cu+Cu, in
Fig.\ \ref{fig3} (c). It is interesting to see that for the
Pratt-radii-ratios between Au+Au and Cu+Cu collisions, all
radii-ratios are flat with $k_T$ and approach $\sim1.4$, which is
equal to the ratio between the initial radii of nuclei.

To summarize, by using the CRAB program, we analyzed the evolution
of the Pratt-radii $R_L$, $R_O$, and $R_S$ at RHIC energies in
collisions simulated by the UrQMD transport model. The calculated
transverse momentum-, centrality-, and system dependence of the
Pratt-radii are in reasonable agreement with the experimental data.
The calculated $R_O$ values for central collisions are $\sim25\%$
larger as compared to experimental data. As a consequence, the extracted quantity
$\sqrt{R_O^{2}-R_S^{2}}$ of the pion emission source is somewhat
larger than experimental estimates.

\section*{Acknowledgments}
We would like to thank S. Pratt for providing the CRAB program and
acknowledge support by the Frankfurt Center for Scientific
Computing (CSC). We thank H.~Appelsh\"auser, T.~J.~Humanic, S. Pratt and M.A. Lisa for valuable
discussions. Q. Li thanks the Alexander von Humboldt-Stiftung for
a fellowship. This work is partly supported by GSI, BMBF, DFG, and
Volkswagenstiftung.

\end{document}